\documentclass[11pt]{article}
\usepackage{times}
\usepackage{geometry}
\usepackage[utf8]{inputenc}
\usepackage{enumitem,amssymb}
\usepackage{ragged2e}
\newlist{thematic}{itemize}{8}
\setlist[thematic]{label=$\square$}
\usepackage{pifont}
\usepackage{color}
\usepackage{graphicx}
\usepackage{wrapfig}
\usepackage{comment}
\usepackage{authblk}

\usepackage[backend=biber, sorting=none, style=nature]{biblatex}
\begin{document}
\begin{flushleft}
\includegraphics[width=2cm]{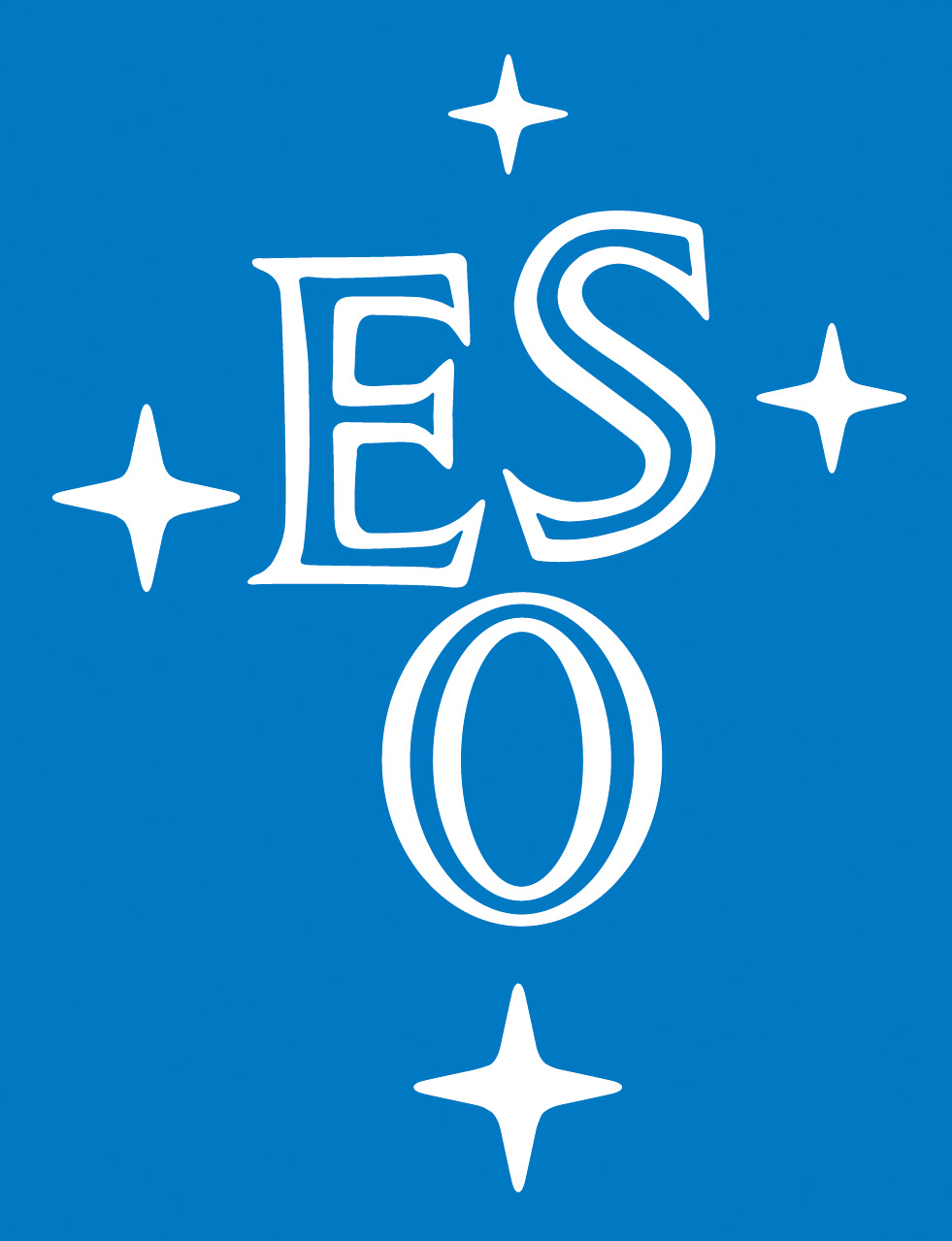} \\[1cm] 
\end{flushleft}
\raggedright
{\Large ESO Expanding Horizon White Paper:}\\ 

\vspace{5mm}
\raggedright
{\LARGE Probing Cosmic Expansion and Early Universe with  Einstein Telescope} \linebreak
\normalsize

\vspace{4mm}

\begingroup
\setlength{\parindent}{0pt}

\textbf{Authors:}\\[1mm]
Angelo Ricciardone$^{1}$,
Mairi Sakellariadou$^{2}$,
Archisman Ghosh$^{3}$,
Alessandro Agapito$^{4}$,
M.~Celeste Artale$^{5}$,
Michael Bacchi$^{6}$,
Tessa Baker$^{7}$,
Marco Baldi$^{8}$,
Nicola Bartolo$^{9}$,
Andrea Begnoni$^{10}$,
Enis Belgacem$^{11}$,
Marek Biesiada$^{12}$,
Jose J.~Blanco-Pillado$^{13}$,
Tomasz Bulik$^{14}$,
Marica Branchesi$^{15}$,
Gianluca Calcagni$^{16}$,
Giulia Capurri$^{1}$,
Carmelita Carbone$^{17}$,
Roberto Casadio$^{8}$,
J.A.R.~Cembranos$^{18}$,
Andrea Cozzumbo$^{15}$,
Ivan De Martino$^{19}$,
Jose M.~Diego$^{20}$,
Emanuela Dimastrogiovanni$^{21}$,
Guillem Domènech$^{22}$,
Ulyana Dupletsa$^{23}$,
Hannah Duval$^{24}$,
Gabriele Franciolini$^{9}$,
Andrea Giusti$^{8}$,
Giuseppe Greco$^{25}$,
Lavinia Heisenberg$^{26}$,
Alexander C.~Jenkins$^{27}$,
Sumit Kumar$^{28}$,
Gaetano Lambiase$^{29}$,
Michele Maggiore$^{11}$,
Michele Mancarella$^{4}$,
Federico Marulli$^{8}$,
Sabino Matarrese$^{9}$,
Isabela Santiago de Matos$^{7}$,
Michele Moresco$^{8}$,
Riccardo Murgia$^{30}$,
Ilia Musco$^{31}$,
Gabriele Perna$^{32}$,
Michele Punturo$^{25}$,
Diego Rubiera-Garcia$^{33}$,
Javier Rubio$^{33}$,
Alexander Sevrin$^{24}$,
Riccardo Sturani$^{34}$,
Matteo Tagliazucchi$^{8}$,
Nicola Tamanini$^{35}$,
Alessandro Tronconi$^{36}$,
Ville Vaskonen$^{37}$,
Daniele Vernieri$^{38}$,
Stoytcho Yazadjiev$^{39}$,
Ivonne Zavala$^{40}$.\\[3mm]

\textbf{Affiliations:}\\[1mm]
$^{1}$ University of Pisa, Pisa, Italy\\
$^{2}$ King’s College London, London, United Kingdom\\
$^{3}$ Ghent University, Ghent, Belgium\\
$^{4}$ Aix-Marseille Université, France\\
$^{5}$ Universidad Andrés Bello, Chile\\
$^{6}$ Universidade Federal do Espírito Santo, Brazil\\
$^{7}$ University of Portsmouth, United Kingdom\\
$^{8}$ University of Bologna, Italy\\
$^{9}$ University of Padua, Italy; INFN and INAF Padova\\
$^{10}$ University of Padua, Italy\\
$^{11}$ Université de Genève, Switzerland\\
$^{12}$ National Centre for Nuclear Research, Poland\\
$^{13}$ IKERBASQUE, Spain; UPV/EHU, Spain\\
$^{14}$ University of Warsaw, Poland\\
$^{15}$ Gran Sasso Science Institute, Italy\\
$^{16}$ IEM--CSIC, Spain\\
$^{17}$ INAF--IASF Milano, Italy\\
$^{18}$ Universidad Complutense de Madrid, Spain\\
$^{19}$ Universidad de Salamanca, Spain\\
$^{20}$ Instituto de Física de Cantabria, Spain\\
$^{21}$ University of Groningen, Netherlands\\
$^{22}$ Leibniz University Hannover, Germany\\
$^{23}$ Austrian Academy of Sciences, MBI, Austria\\
$^{24}$ Vrije Universiteit Brussel, Belgium\\
$^{25}$ Istituto Nazionale di Fisica Nucleare (INFN), Sezione di Perugia, Italy\\
$^{26}$ Heidelberg University, Germany\\
$^{27}$ University of Cambridge, United Kingdom\\
$^{28}$ Utrecht University, Netherlands\\
$^{29}$ University of Salerno, Italy\\
$^{30}$ University of Cagliari, Italy\\
$^{31}$ University of Nova Gorica, Slovenia\\
$^{32}$ KBFI, Tallinn, Estonia\\
$^{33}$ IPARCOS--UCM, Universidad Complutense de Madrid, Spain\\
$^{34}$ Instituto de Física Teórica, UNESP, Brazil\\
$^{35}$ Laboratoire de 2 Infinis, Toulouse, France\\
$^{36}$ INFN Sezione di Bologna, Italy\\
$^{37}$ KBFI, Estonia\\
$^{38}$ University of Naples, Italy\\
$^{39}$ Sofia University, Bulgaria\\
$^{40}$ Swansea University, United Kingdom\\[3mm]

\textbf{Contact emails:} \texttt{angelo.ricciardone@unipi.it}; 

\endgroup

\pagebreak
\justifying
\section{Cosmology with Einstein Telescope}

Over the next two decades, gravitational-wave (GW) observations are expected to evolve from a discovery-driven field into a precision tool for astrophysics and cosmology.  The achievements of current second-generation detectors (LIGO-Virgo and KAGRA)~\cite{LIGOScientific:2016aoc} have established the existence of compact-binary mergers and opened the GW multi-messenger astronomy~\cite{LIGOScientific:2017vwq}. Despite these breakthroughs, present observatories remain limited in sensitivity, redshift reach and duty cycle, preventing them from addressing many of the most fundamental open questions in cosmology and gravity.
As we approach the 2040s, electromagnetic facilities such as DESI~\cite{DESI:2023dwi}, LSST/Rubin~\cite{LSSTScience:2009jmu} and Euclid~\cite{EUCLID:2011zbd} (already operating) and future ones like 4MOST~\cite{2012SPIE.8446E..0TD} and
MUltiplexed Survey Telescope (MUST)~\cite{Zhao:2024alp}, will have mapped the luminous Universe with deep precision. Nevertheless, several key problems are likely to remain unresolved. The nature of dark matter, the physical origin of the cosmic acceleration, the physical conditions of the first fractions of seconds after the Big Bang, and the behaviour of gravity across cosmological distances cannot be fully constrained with electromagnetic tracers alone. A possible way to answer these questions will be to benefit from accessing physical processes that do not emit light. GWs provide precisely such a channel: they carry information from epochs and scales where electromagnetic approaches lose sensitivity, travel cosmological distances almost unaffected by intervening matter, and respond directly to the geometry of spacetime. In this context, next generation GW observatories, such as the Einstein Telescope (ET), emerges as an indispensable facility for European astronomy. ET is conceived to operate in a frequency band and at a sensitivity level beyond the capabilities of existing ground-based detectors. This will enable access to binary black holes and neutron stars out to redshifts unreachable today, provide continuous monitoring of sources with high signal-to-noise ratios, and allow to detect gravitational-wave backgrounds (GWBs) of astrophysical and cosmological origin. Its capabilities will complement and extend those of space-based detectors, such as LISA~\cite{LISA:2024hlh}. While LISA will explore the Galaxy and massive black-hole mergers, ET will probe the rapid dynamical environments of neutron stars, stellar-origin black holes, and the fundamental processes governing the cosmic evolution.

\subsection{Can gravitational-waves reveal what happened in the earliest stages of the Universe?}
Broader frequency coverage and one-order of magnitude improved sensitivity will allow to open a new observational window on the primordial Universe, where tiny fluctuations in spacetime generated by high-energy physics are inaccessible by current detectors. A major scientific opportunity lies in accessing the physics of the early Universe.  
Unlike photons, which were decoupled 380,000 years after the Big Bang, primordial GWs are relics of the Planck epoch. 
Sensitive GW measurements will allow the exploration of a wide range of GWB sources, probing 
inflationary models beyond the standard slow-roll paradigm, including scenarios with gauge fields, axion dynamics, or non-minimal gravitational couplings. These observations will also make it possible to probe first-order phase transitions in the early Universe, which generate powerful bursts of GWs through bubble collisions and plasma turbulence. For phase transitions occurring at temperatures up to $10^3$\,GeV, GW data taken by next-generation observatories will provide the only means of detecting the corresponding GWB signal. Networks of cosmic strings, if produced at any stage in the early Universe, would emit GWs across more than twenty orders of magnitude in frequency. High-precision GW measurements, that are not achievable with current detectors, will provide the opportunity to reconstruct the spectral shape, measure the string tension down to $G\mu\simeq 10^{-19}$, and test grand unified theories. Similarly, more sensitive GW observations will provide access to scalar-induced GWs associated with the formation of primordial black holes, offering a direct probe of inflationary small-scale perturbations and possible candidates for dark matter~(see \cite{ET:2025xjr} and reference therein).

\subsection{Can we shed light on the Hubble tension with gravitational waves?}

Cosmography with GWs will become a true precision tool as detector sensitivity increases, because compact binary mergers provide an absolute measurement of the luminosity distance. Next generation observatories will detect them in unprecedented numbers and across an extended redshift range. Current detectors are limited to relatively small samples of standard sirens, but future observatories, such as ET, are expected to detect on the order of $\mathcal{O}(10^5)$ BNS mergers per year, with realistic follow-up strategies yielding roughly $10^3$ bright standard sirens over a decade up to $z\sim 0.5$, and an additional $\sim 10^2$ short-GRB associations extending to $z\sim1$, considering ET in combination with THESEUS~\cite{ET:2025xjr,theseus:2021}. These bright-siren samples alone will allow percent-level determinations of $H_0$ and $\Omega_{m,0}$, and improvements of more than an order of magnitude on curvature constraints when combined with forthcoming large-scale structure surveys. With an observatory, such as ET,
dark-siren cosmography exploits 
$\mathcal{O}(10^5)$ BBH detections per year, with sources that can be as distant as redshift $z\gtrsim 10$. Statistical association with deep galaxy catalogs—expected to contain $10^9$–$10^{10}$ galaxies from surveys such as LSST, Euclid, DESI and WST—enables sub-percent $H_0$ precision within a single year when ET is part of a 3G network, provided spectroscopic coverage is sufficient for the best-localized events.  
Spectral-siren approaches, which use identifiable mass-scale features in NS and BH populations, provide an independent handle on redshift by breaking the mass–redshift degeneracy at the population level, and become especially powerful when informed by population-synthesis–based merger-rate models.  
With ET alone, tidal deformability measurements allow redshift determinations with $\sim 10\%$ precision up to $z\sim1$, while a joint ET+CE network can improve this to $\sim1\%$. For nearby systems ($d\lesssim 100$ Mpc) with detectable post-merger spectra, redshifts may be recovered at the $\sim1\%$ level, enabling $H_0$ measurements approaching $0.1\%$ precision when combined with high-SNR inspiral detections~\cite{Branchesi:2023mws}.  Taken together, these methods show that GW observations by next-generation detectors will deliver a high-precision GW Hubble diagram from the local Universe to beyond the peak of cosmic star formation. Quantitatively, ET cosmography is expected to achieve sub-percent measurements of $H_0$, and robust, model-independent reconstructions of the distance–redshift relation, shedding light on the so-called Hubble tension~\cite{DiValentino:2021izs}.

\subsection{Do gravitational waves and light measure the same cosmic expansion, or do gravitational waves reveal deviations from General Relativity?  }

In the standard $\Lambda$CDM model, cosmic acceleration is encoded in a dark-energy component with equation of state $w_{\rm DE}(z)$ and a corresponding EM luminosity distance $d_L^{\rm em}(z)$, which has been already tested at few-percent level by CMB, BAO and SNe. In contrast, the tensor sector is still poorly constrained: in many modified-gravity models the GW propagation equation acquires a modified friction term, so that compact binaries do not measure $d_L^{\rm em}(z)$ but a GW luminosity distance $d_L^{\rm gw}(z)\neq d_L^{\rm em}(z)$. 
Current limits from GW170817, dark sirens and population analyses are still dominated by large uncertainties
leaving ample room for scenarios in which the background and scalar sectors agree with $\Lambda$CDM at the few-percent level while the tensor sector shows large deviations as large in $d_L^{\rm gw}$. GW observations by next-generation observatories will change this picture by providing thousands of standard sirens with spectroscopic redshifts from ESO facilities and other EM surveys, enabling percent-level measurements of the $d_L^{\rm gw}/d_L^{\rm em}$ relation~\cite{Branchesi:2023mws}. 
GW observations by next-generation observatories, in synergy with LSS probes, are expected to test if the GW Hubble diagram agree with the EM one, and to rule out modified-gravity models that perfectly mimic $\Lambda$CDM in the background while leaving a large imprint in the GW propagation.

\subsection{How does the large-scale structure map the gravitational wave sky?}

Achieving sensitivity sufficient to detect essentially all compact binary mergers, or a large fraction of it,
will provide a new, high-redshift tracer of the large-scale structure. Unlike galaxies, GW sources populate luminosity-distance space, and their cross-correlation with redshift-space tracers such as Euclid or DESI galaxies carries cosmological information with different parameter degeneracies. Forecasts show that combining ten years of ET observations with two Cosmic Explorer (CE) detectors and a Euclid-like photometric survey (20 tomographic bins up to $z\simeq 2$) enables a sub-percent measurement of the Hubble constant, with $\sigma(H_0)\simeq 0.5$–$0.8\%$, and a precise determination of the matter density, $\sigma(\Omega_m)\simeq 1\%$, after marginalising over more than 40 nuisance parameters. These constraints rely only on GW–galaxy auto- and cross-correlations and do not assume external CMB or BBN priors. Bright sirens add additional constraining power by tracing both density and velocity fields. Joint analyses of ET BNSs with kilonova counterparts (up to $z\sim 0.5$), Rubin Observatory imaging ($\sim 10^4\,\mathrm{deg}^2$), and DESI galaxies show that a full $6\times2$-point analysis leads to striking gains: the precision on $H_0$ improves by $\sim 30\%$ relative to distance-only methods, while curvature constraints sharpen by more than an order of magnitude, reaching $\sigma(\Omega_{k0})<0.02$ within five years. Constraints on the growth of structure also become competitive, with uncertainties on $\sigma_8$ and the growth-rate index $\gamma$ approaching percent levels. ET will also access the anisotropies of the astrophysical GWB, whose clustering bias encodes the environments of compact binaries. Even detecting only the first few multipoles in AGWB–galaxy or AGWB–CMB cross-correlations can constrain the GW bias at the $\sim$5–10\% level, sufficient to distinguish astrophysical black-hole populations from primordial black holes, thereby probing their possible contribution to dark matter. Beyond cross-correlation with electromagnetic surveys, ET+CE will be capable of reconstructing the LSS directly from GW data alone. With thousands of BBHs localised to $\Delta\Omega\lesssim 1\,\mathrm{deg}^2$ per year, 3G networks can recover the BBH clustering bias within 5–10 years and detect the BAO feature at $\sim 100\,h^{-1}\mathrm{Mpc}$ purely from GW clustering. This yields an independent measurement of $H_0$ and $\Omega_m$ competitive with galaxy surveys. 

In summary, GW detections from next-generation observatories can address long-standing questions in cosmology that remain inaccessible with current detectors, whose limited sensitivity and frequency coverage prevent exploration of key regimes. In addition, GW observations  will make it possible to probe sources and physical processes unreachable with electromagnetic observations alone. The European next-generation observatory, ET stands at the forefront of this vision. By extending the observable frequency range and achieving sensitivities far beyond those of second-generation detectors, ET will detect compact-object mergers—binary black holes and binary neutron stars—across the full extent of the observable Universe. Moreover, its unprecedented sensitivity to GWBs will enable the study of cosmological and astrophysical processes that have so far remained entirely out of reach.
\section*{References}
[1] Abbott, B. P. et al. Observation of Gravitational Waves from a Binary Black Hole Merger. Phys. Rev. Lett. 116, 061102;
[2] Abbott, B. P. et al. GW170817: Observation of Gravitational Waves from a Binary Neutron Star Inspi- ral. Phys. Rev. Lett. 119, 161101;
[3] Adame, A. G. et al. Validation of the Scientific Program for the Dark Energy Spectroscopic Instrument. Astron. J. 167, 62;
[4] Abell, P. A. et al. LSST Science Book, Version 2.0. arXiv: 0912.0201 [astro-ph.IM]; 
[5] Laureijs, R. et al. Euclid Definition Study Report. arXiv: 1110.3193 [astro-ph.CO];
[6] 4MOST: 4-metre multi-object spectroscopic telescope in Ground-based and Airborne Instrumentation for Astronomy IV (eds McLean, I. S., Ramsay, S. K. \& Takami, H.) 8446 (Sept. 2012), 84460T. arXiv: 1206.6885 [astro-ph.IM];
[7] Zhao, C. et al. MUltiplexed Survey Telescope: Perspectives for Large-Scale Structure Cosmology in the Era of Stage-V Spectroscopic Survey. arXiv: 2411.07970 [astro-ph.CO];
[8] Colpi, M. et al. LISA Definition Study Report. arXiv: 2402.07571 [astro-ph.CO];
[9] Abac, A. et al. The Science of the Einstein Telescope. arXiv: 2503.12263 [gr-qc]; 
[10] Amati, L. et al. The THESEUS space mission: science goals, requirements and mission concept. Experimental Astronomy 52, 183–218;
[11] Branchesi, M. et al. Science with the Einstein Telescope: a comparison of different designs. JCAP 07,
068;
[12] Di Valentino, E. et al. In the realm of the Hubble tension—a review of solutions. Class. Quant. Grav.
38, 153001.

\end{document}